\documentclass[aip,jcp,reprint]{revtex4-1}

\usepackage{graphicx}
\usepackage{dcolumn}
\usepackage{bm}
\usepackage[utf8]{inputenc}
\usepackage[T1]{fontenc}
\usepackage{mathptmx}
\usepackage{etoolbox}
\usepackage{balance}
\usepackage{soul}
\usepackage{graphicx}

\usepackage{physics2}
\usepackage{xcolor, soul}
\usepackage{multirow, tabularx, booktabs}
\usepackage{braket}
\usepackage{adjustbox}
\usepackage{tablefootnote}

\usepackage{xr} 
\usepackage{threeparttable}
\usephysicsmodule{braket}

\usepackage[version=4]{mhchem}

\usepackage{physics}
\usepackage{amsmath}

\newcommand{\Tp}{{\hat{T}_{\rm pCCD}}}   
\newcommand{\pccd}{{\ket{\rm pCCD}}}

\newcommand{\Lower}[1]{\smash{\lower 1.5ex \hbox{#1}}}
\makeatletter
\def\@email#1#2{%
 \endgroup
 \patchcmd{\titleblock@produce}
  {\frontmatter@RRAPformat}
  {\frontmatter@RRAPformat{\produce@RRAP{*#1\href{mailto:#2}{#2}}}\frontmatter@RRAPformat}
  {}{}
}
\makeatother

\begin{document}

\preprint{AIP/123-QED}
\title[Orbital energies and pair-orbital energies from pCCD-based methods]
{Simple and efficient computational strategies for calculating orbital energies and pair-orbital energies from pCCD-based methods}
\author{Seyedehdelaram Jahani}
\affiliation{Institute of Physics, Faculty of Physics, Astronomy, and Informatics, Nicolaus Copernicus University in Toruń, Grudziądzka 5, 87-100 Toru\'{n}, Poland.}
\author{Somayeh Ahmadkhani}
\affiliation{Institute of Physics, Faculty of Physics, Astronomy, and Informatics, Nicolaus Copernicus University in Toruń, Grudziądzka 5, 87-100 Toru\'{n}, Poland.}
\author{Katharina Boguslawski}
\affiliation{Institute of Physics, Faculty of Physics, Astronomy, and Informatics, Nicolaus Copernicus University in Toruń, Grudziądzka 5, 87-100 Toru\'{n}, Poland.}
\author{Pawe\l{} Tecmer}
\affiliation{Institute of Physics, Faculty of Physics, Astronomy, and Informatics, Nicolaus Copernicus University in Toruń, Grudziądzka 5, 87-100 Toru\'{n}, Poland.}
 \email{ptecmer@fizyka.umk.pl}

\date{\today}
\begin{abstract}
We introduce affordable computational strategies for calculating orbital and pair-orbital energies in atomic and molecular systems. 
Our methods are based on the pair Coupled Cluster Doubles (pCCD) ansatz and its orbital-optimized variant. 
The computed orbital and pair-orbital energies are then subsequently used to approximate ionization potentials (IP), electron affinities (EA), the resulting charge gaps, double ionization potentials (DIP), and double electron affinities (DEA). 
Our methodology builds on the standard Koopmans' theorem and {refnes} it for a pCCD-based wave function. 
Furthermore, we incorporate pCCD electron correlation effects into the model utilizing canonical Hartree-Fock or natural pCCD-optimized orbitals.
The latter represents a diagonal approximation to the (D)IP/D(EA) equation of motion pCCD models.
We benchmarked our newly developed models against theoretical and available experimental data for selected atoms in various basis set sizes and a set of 24 organic acceptor molecules.
Our numerical results show that the Koopmans' approach based on pCCD natural orbitals provides a balanced treatment of occupied and virtual orbital energies, resulting in reliable predictions of charge gaps at a low computational cost.
\end{abstract}

\maketitle
\begin{quotation}
\end{quotation}
\section{\label{sec:introduction}Introduction}

During the past few decades, the field of organic electronics~\cite{takimiya2014pi} focused on studying a wide range of materials, not only for their academic interest but also for applications in organic solar cell development.~\cite{su2012organic}
The photovoltage of organic-based materials continuously improved over the past years, and their power conversion efficiency can now reach over 19$\%$ in the laboratory.~\cite{zhu2022single}
Low environmental impact and production costs of organic photovoltaics (OPVs)~\cite{zhu2022single} materials allow them to compete with silicon-based inorganic alternatives. 
An advantage of OPV-based materials is the variety of organic molecules that can be utilized as building blocks of donors and acceptors.
The energy difference between the Highest Occupied Molecular Orbital (HOMO) of the donor and the Lowest Unoccupied Molecular Orbital (LUMO) of the acceptor (the HOMO-LUMO gap) is one of the initial tuning parameters to steer opto-physical properties. 
Consequently, it plays a vital role in evaluating a molecule's reactivity and stability, influencing its behavior in various chemical processes and applications.
To that end, a reliable prediction of the HOMO-LUMO gaps is indispensable in screening new building blocks and doping strategies of organic molecules. 

While methods like Hartree--Fock theory and Density Functional Approximations (DFAs)~\cite{parr1979local, cohen2012challenges} can provide this information at low computational cost by utilizing the Koopmans' theorem~\cite{koopmans-theorem-1934} and Janaks' theorem~\cite{janak_theorem}, respectively, they do not always yield reliable results.~\cite{zhang2007comparison, körzdörfer2014organic}
Hartree--Fock theory does not incorporate electron correlation effects (except of exchange correlation), and the virtual orbitals often lack the correct physical meaning and poorly describe LUMOs. 
Specifically, the aforementioned electronic structure methods are frequently inadequate for reliably describing sizable $\pi$-extended systems,~\cite{mainville2021theoretical, tortorella2016benchmarking, körzdörfer2014organic} which form crucial constituents of OPVs.~\cite{takimiya2014pi}
Typical deficiencies and drawbacks include the absence of systematic improvability, delocalization errors, and the inadequacy to describe static electron correlation effects.~\cite{cohen2012challenges}
For instance, semi-local hybrid exchange--correlation functionals often overestimate the dispersion of electron and hole densities, while range-separated hybrids tend to confine electron densities within $\pi$-conjugated chains overly.~\cite{cai2002failure, cohen2008fractional, mori2008localization}
These challenges highlight the limitations of current computational methods to reliably and efficiently predict fundamental electronic properties of organic molecules.~\cite{dreuw2004failure}

In contrast, reliable quantum chemistry methods are often too expensive for large-scale modeling of $\pi$-conjugated systems and the building blocks of organic photovoltaics (OPVs). Unconventional approaches, such as geminal-based methods~\cite{parr_1956, bardeen_1957, parks_1958, coleman_1965, miller1968, surjan_1999, gvb-cc-jctc-2014, block-corrected-cc} provide more cost-effective and intuitive means of modeling electronic structures and properties of such systems.~\cite{pawel-pccp-geminal-review-2022, pccd-perspective-jpcl-2023}
These methods represent a promising alternative to traditional quantum chemistry techniques by integrating pair-wise electron correlation schemes into electronic wave functions.~\cite{gvb-cc-jctc-2014, pernal2014, pernal2014_jcp_erratum, li2007generalized, pawel-pccp-geminal-review-2022, johnson_2013, johnson2017strategies, johnson2020richardson, johnson2022bivariational, faribault2022, fecteau2022, moisset2022}
One promising geminal model is the pair Coupled Cluster Doubles (pCCD) ansatz,~\cite{boguslawski2014efficient, tamar-pccd} also known as the Antisymmetric Product of 1-reference orbital Geminal (AP1roG) ansatz.~\cite{limacher_2013}
pCCD-based numerical examples include the one-dimensional Hubbard model, ~\cite{boguslawski2014efficient, tamar-pccd, ps2-ap1rog, boguslawski2014nonvariational} 
molecules with stretched bonds,~\cite{tamar-pccd, pawel_jpca_2014, ps2-ap1rog, frozen-pccd, kasia-orbital-entanglement-ijqc-2015, ijqc-eratum, pawel-pccp-2015, garza2015actinide, ap1rog-lcc, pccd-prb-2016, pccd-ci, boguslawski2017benchmark, filip-jctc-2019, state-specific-oopccd, pawel-yb2, ola-tcc} heavy-elements containing compounds, ~\cite{pawel-pccp-2015,garza2015actinide,pccd-ee-f0-actinides, pawel-yb2, ola-qit-actinides-pccp-2022, galynska2024mocofactor} dipole moments~\cite{pccd-dipole-moments-jctc-2024} and embedding.~\cite{pccd-static-embedding} 
Specifically, pCCD combined with an orbital optimization protocol can be a promising alternative to DFAs for modeling organic electronics.~\cite{pccd-delaram-rsc-adv-2023, pccd-perspective-jpcl-2023, galyńska2024benchmarking}

However, wave function-based methods, including geminal-based approaches, often do not provide information or computationally affordable prescriptions for obtaining orbital energies (in contrast to (natural) occupation numbers).
To address this issue, we propose and investigate various models for determining orbital energies from pCCD-based wave functions, focusing on scalable quantum chemical approaches that are applicable to realistic OPV materials. 
Our computational strategies are inspired by the work of Limacher,~\cite{piotrus-orbital-energies}, who exploited geminal-based pair-orbital energies for subsequent perturbation theory corrections. 
In this work, we aim to obtain computationally efficient methods for calculating orbital and pair-orbital energies from which approximate ionization potentials (IP), electron affinities (EA), double ionization potentials (DIP), and double electron affinities (DEA) {and compare them to the existing (D)IP-EOM-pCCD and (D)EA-EOM-pCCD implementations.
The accuracy of these pCCD-EOM-based (D)IPs and EAs has been extensively benchmarked, including the commonly used exchange--correlation functionals. \mbox{~\cite{mamache2023benchmarking, galyńska2024benchmarking,galyńska2024exploring}}
}

{Specifically, we introduce the modified Koopmans' theorem, which improves upon the standard Koopman's approach, by incorporating some fraction of electron correlation effects into the model.
Theoretical foundations for that model are provided in section \mbox{~\ref{sec:theory}} (vide infra).
That approach differs from the well-known Extended Koopmans' Theorem (EKT), representing the generalization of Koopmans' theorem to correlated wave functions.\mbox{~\cite{bozkaya2014accurate, smith1975extension, day1975extension}} 
Opposite to EKT, our modified Koopman's approach does not require diagonalization of the generalized Fock matrix to compute orbital energies (used to approximate IPs and EAs).

It is important to stress that Koopman's-based approaches are not the only way to compute ionization potentials at a reduced cost. 
Other examples are, for instance, the semi-canonical second-order optimized effective potential (OEP2-sc) method and its scaled-opposite-spin (SOS) variant (OEP2-SOS-sc),
\mbox{~\cite{Siecinska2022}} the low-scaling equation-of-motion coupled-cluster theory with single and double (EOM-CCSD) excitations and its second-order many-body perturbation theory (EOM-MBPT(2)) approximations,\mbox{~\cite{park2018low}} the low-scaling Green's Wave function (GW) algorithms,\mbox{~\cite{wilhelm2021low}} 
the Møller--Plesset energy differences and its SOS variant ($\Delta$MP2-SOS(IP)),\mbox{~\cite{smiga2018spin}} and 
the second-order algebraic diagrammatic construction (ADC(2))\mbox{~\cite{shaalan2022accurate}}.}
This work is organized as follows. Section~\ref{sec:Theory} provides a brief review of the investigated theoretical models, followed by an explanation of the computational methodology in section~\ref{sec:comput-det}.
Section~\ref{sec:results} presents a summary of numerical results and statistical analysis, leading to concluding remarks in section~\ref{sec:conclusions}.

\section{Theory}\label{sec:Theory}
\subsection{pCCD}
The pCCD wave function ansatz~\cite{limacher_2013, tamar-pccd, pawel-pccp-geminal-review-2022,boguslawski2014efficient} can be expressed as a simplification of the CCD model,
\begin{equation}
\label{eqn:equation-pccd}
     \pccd = e^{\Tp} \ket{\phi_0},
\end{equation}   
where
\begin{equation}
\label{eqn:equation-tpccd}
     \Tp = \sum_{i=1}^{n_{\rm occ}} \sum_{a=1}^{n_{\rm virt}} 
    t^{a\bar{a}}_{i\bar{i}} a^\dagger_a a^\dagger_{\bar{a}} a_{\bar{i}} a_i
\end{equation}
is a cluster operator that excites electron pairs, ${ \hat{a}_{p} ({\hat{a}^{\dagger}_{\bar{p}}})  }$ are the elementary annihilation (creation) operators for spin-up ${p}$ and spin-down ${(\bar{p})}$ electrons, {${t^{a\bar{a}}_{i\bar{i}}}$} are the pCCD cluster amplitudes, and $\ket{\phi_0}$ is some independent-particle wave function, for example, the Hartree--Fock (HF) determinant.
The above sum runs over all occupied ${i}$ and virtual ${a}$ orbitals.
One advantage of using the exponential form is the proper (linear) scaling of the method with the particle number (size extensivity), while size consistency is guaranteed by applying an orbital optimization protocol.~\cite{boguslawski2014efficient, piotrus_mol-phys, tamar-pccd, ps2-ap1rog, boguslawski2014nonvariational, state-specific-oopccd}
\subsection{Orbital energies from Koopmans' and modified Koopmans' theorem}
\label{sec:theory}
In the following, we connect the orbital and pair-orbital energies based on Koopmans' theorem and introduce simple correction to incorporate electron correlation effects from pCCD. The latter will be referred to as a modified Koopmans's theorem.  
\subsubsection{Koopmans' theorem}
Koopmans' theorem is the simplest way to give orbital energies a physical meaning.~\cite{koopmans-theorem-1934, szabo_book}
According to the theorem, which was originally derived for the canonical Hartree--Fock orbitals, the negative value of the occupied orbitals' energy approximates ionization potentials (IP), and the negative one of the virtual (unoccupied) orbitals' energies, to a large extent, estimates electron affinities (EA).
Specifically, the energy differences when adding or removing one electron are expressed as~\cite{koopmans-theorem-1934}
\begin{equation}
\label{eqn:ip}
      E_{N-1} - E_N = {\langle 0|{a^\dagger_i} \hat{H} {a_i}| 0\rangle - \langle0|\hat{H}|0 \rangle = -{f}_{ii} = -\epsilon_i  }
\end{equation}
for the removal of an electron and
\begin{equation}
\label{eqn:ea}
    E_{N} - E_{N+1} = \langle0|\hat{H}|0 \rangle - {\langle 0|{a_a} \hat{H} {a^\dagger_a}| 0\rangle = - {f}_{aa} = -\epsilon_a}
\end{equation}
for the attachment of one.
In the above equations, $f_{ii}$ and $f_{aa}$ are diagonal elements of the Fock matrix corresponding to occupied $\epsilon_i$ and virtual $\epsilon_a$ orbital energies.
The $f_{pq}$ element of the inactive Fock matrix is defined as (within the restricted orbital picture)
\begin{equation}
      \label{eqn:fock}
        f_{pq} = h_{p q}+\sum_i^{\mathrm{occ}} ( 2 \bra{p i}\ket{q i} - \bra{p i}\ket{i q}), 
\end{equation}
where ${h_{pq}}$ are the one-electron integrals, $\bra{p i}\ket{q i}$ and $\bra{p i}\ket{i q}$ are the two-electron (Coulomb and exchange) integrals in physicist' notation.
Throughout this work, we will restrict all expressions to the case of restricted orbitals, where the spatial orbitals are equivalent for the $\alpha$ and $\beta$ spin components and the number of occupied orbitals is equal to the number of (active) electron pairs $n_{\rm nocc} = N_{\rm el}/2$.
In Eqs.~\eqref{eqn:ip} and \eqref{eqn:ea}, we use the molecular electronic Hamiltonian in its second-quantized form,
\begin{equation}
      \label{eqn:equation-hamiltonian}
    {{\hat{H}} ={\sum_{pq}^{k}}{h_{pq}}{\sum_{\sigma}}  {\hat{a}^{\dagger}_{{p}{\sigma}}} {{\hat{a}_{{q}{\sigma}}}}+ {\frac{1}{2}} {\sum_{pqrs}^{k}} {\langle {pq|rs}\rangle}{\sum_{{\sigma}{\bar{\sigma}}}} {{\hat{a}^{\dagger}_{{p}{\sigma}}}} {{\hat{a}^{\dagger}_{{q}{\bar{\sigma}}}}}          {{\hat{a}_{{s}{\bar{\sigma}}}}} {{\hat{a}_{{r}{\sigma}}}} }
\end{equation}
In the above equation, the indices ${p}$, ${q}$,... indicate all occupied and virtual orbitals, while ${\sigma}$ (${\bar{\sigma}}$) encodes the spin degree of freedom (${ \alpha}$ (${{\beta}}$)).

Analogously, when two electrons are added to or removed from some orbital pair $i,j$, the corresponding energy differences are given by~\cite{piotrus-orbital-energies, szabo_book}
\begin{align}
\label{eqn:dip}
      E_{N-2} - E_{N} &= \langle 0|{a}^\dagger_i{a}^\dagger_{j} \hat{H} {a}_{j}{a_i}| 0\rangle - \langle0|\hat{H}|0 \rangle = -{f}_{ii} -{f}_{jj} + V_{ijij} 
\end{align}
and
\begin{align}
\label{eqn:dea}
    E_{N} - E_{N+2} &= \langle 0|\hat{H}|0 \rangle - \langle 0|{a}_{{b}}{a_a} \hat{H} {a}^\dagger_a {a}^\dagger_{b}| 0\rangle = - {f}_{aa} - {f}_{bb} - V_{abab}
\end{align}
respectively. In the above equations, $V_{pqpq} = \langle{pq \| pq\rangle} + \langle{p\bar{q} | p\bar{q}\rangle}$ contains both Coulomb and exchange terms.
Depending on the spin of the electron occupying orbital $p$, either the Coulomb ($p\bar{q}$ case) or both Coulomb and exchange terms ($pq$ case) are included.

According to Koopmans' theorem, the energy corresponding to occupied and virtual orbitals is deduced from the IPs (see eq.~\eqref{eqn:ip}) and EAs (see eq.~\eqref{eqn:ea}), respectively.
Thus, the relations for orbital ($\varepsilon_{p}$) and pair-orbital energies ($\varepsilon_{pq}$) are as follows,
\begin{equation}
      \label{eqn:ep}
 {\varepsilon_{{p}}} = f_{p},
\end{equation}
\begin{equation}
    \label{eqn:eij}
    \varepsilon_{ij} = f_{ii} + f_{jj} - \left\langle ij \middle\| ij \right\rangle
    \quad \mathrm{and} \quad
    \varepsilon_{i\bar{j}} = f_{ii} + f_{\bar{j}\bar{j}} - \left\langle i\bar{j} \middle| i\bar{j} \right\rangle
\end{equation}
\begin{equation}
      \label{eqn:eab}
      \varepsilon_{ab} = f_{aa} + f_{bb} + \left\langle ab \middle\| ab \right\rangle
      \quad \mathrm{and} \quad
      \varepsilon_{a\bar{b}} = f_{aa} + f_{\bar{b}\bar{b}} + \left\langle a\bar{b} \middle| a\bar{b} \right\rangle,
\end{equation}
where we distinguish between same- and opposite-spin cases, $\varepsilon_{pq}$ and $\varepsilon_{p\bar{q}}$, respectively.
Thus, $\varepsilon_{pq}$ and $\varepsilon_{p\bar{q}}$ approximate the triplet ($pq$) and singlet ($p\bar{q}$) sector for DIP \eqref{eqn:dip} and DEA \eqref{eqn:dea} processes, respectively.
Here, we enforce $\varepsilon_{pp} = 0$ to avoid the removal/addition of an electron from/to the same orbital twice.

Koopmans' theorem is generally well-defined for a canonical HF reference function within the single-determinant formalism.
Here, we aim to extend Koopmans' theorem to a non-canonical reference function and combine it with a pCCD reference state, moving to a multi-determinant picture.
By doing so, we arrive at different levels of approximations to extract orbital energies from the modified expressions of IPs and EAs and pair-orbital energies from the modified equations of DIPs and DEAs, respectively.
In the first approximation, we evaluate eqs.~\eqref{eqn:ip}, \eqref{eqn:ea}, \eqref{eqn:dip}, and \eqref{eqn:dea} for the reference determinant of an orbital-optimized (oo)-pCCD state,~\cite{boguslawski2014efficient, boguslawski2014nonvariational}
indicated by $\ket{0_\textrm{pCCD}}$,

\begin{align}
\label{eqn:ip-pccdre}
    \textrm{IP}_{0_{\rm pCCD}} &= E_{N-1} - E_N \notag\\
    &= \langle 0_{\rm pCCD}|{a^\dagger_i} \hat{H} {a_i}| 0_{\rm pCCD}\rangle \nonumber - \langle0_{\rm pCCD}|\hat{H}|0_{\rm pCCD} \rangle \nonumber \notag\\
    &= -{f}_{ii}, 
\end{align}

and
\begin{align}
\label{eqn:ea-pccdre}
            \textrm{EA}_{0_{\rm pCCD}} &= E_{N} - E_{N+1} \notag\\ &= \langle 0_{\rm pCCD}|{a^\dagger_a} \hat{H} {a_a}| 0_{\rm pCCD}\rangle - \langle0_{\rm pCCD}|\hat{H}|0_{\rm pCCD} \rangle  \nonumber \\
            &= -{f}_{aa} . 
\end{align}
Since we only changed the reference determinant, the final expressions formally remain the same, except that all one- and two-electron integrals are determined for the oo-pCCD natural orbitals.
The same holds for DIP and DEA in eqs.~\eqref{eqn:dip} and \eqref{eqn:dea}.

\subsubsection{Modified Koopmans' theorem}
\begin{figure*}[tb]
\centering
\includegraphics[scale=1.0]{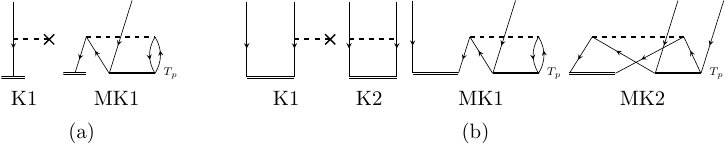}
\caption{Diagrammatic representation of the energy expressions of the (a) IP and (b) DIP equations using antisymmetrized Goldstone diagrams.
K labels terms that appear in the conventional Koopmans' theorem. MK indicates each correction term emerging from using $\hat{H}_N^{(\rm pCCD)}$ in the IP/EA expressions of the modified Koopmans' theorem.
Open lines indicate the $\bra{0} a^\dagger_i \ldots $ bra states.
The double lines label the $ \ldots a_i \ket{0}  $ ket states and are used to highlight the relation to the (D)IP-EOM-pCCD models.
The corresponding (D)EA diagrams can be obtained by reversing the direction of all open lines.
}
\label{fig:diagrams}
\end{figure*}

A first correction term is obtained by changing the Hamiltonian for which eqs.~\eqref{eqn:ip}, \eqref{eqn:ea}, \eqref{eqn:dip}, and \eqref{eqn:dea} are evaluated.
By doing so, we move from the single- to the multi-determinant picture, incorporating correlation effects from the pCCD state.
Replacing $\hat{H}$ by the similarity-transformed Hamiltonian of pCCD, $\hat{H}^{(\rm pCCD)} = e^{-\Tp} \hat{H} e^\Tp$, the modified IP expression involving orbital $i$, then, reads
{\small
\begin{align}
\label{eqn:ip-pccdref}
E_{N-1} - E_N &= \langle 0 | a_i^\dagger \hat{H}^{(\rm pCCD)} a_i | 0 \rangle - \langle 0 | \hat{H}^{(\rm pCCD)} | 0 \rangle \notag\\
&= \langle 0 | a_i^\dagger (\hat{H}^{(\rm pCCD)} - E_{\text{ref}}) a_i | 0 \rangle - \langle 0 | (\hat{H}^{(\rm pCCD)} - E_{\text{ref}}) | 0 \rangle \notag\\
&= \langle 0 | a_i^\dagger \hat{H}^{(\rm pCCD)} _N a_i | 0 \rangle 
- \langle 0 | \hat{H}^{(\rm pCCD)} _N | 0 \rangle \notag\\
&= \langle 0 | a_i^\dagger \left[ \hat{H}^{(\rm pCCD)} _N , a_i \right] | 0 \rangle \notag\\
&= -f_{ii} - \sum_{c} t^{c\bar{c}}_{i\bar{i} } \langle i\bar{i} |c\bar{c} \rangle,
\end{align}
}
where we introduced the normal-product form of the Hamiltonian, $\hat{H}_N^{(\rm pCCD)} = \hat{H}^{(\rm pCCD)} - \langle 0 | \hat{H}| 0 \rangle$.
Analogously, we obtain for the electron attachment energies related to orbital $a$,
\begin{align}
\label{eqn:ea-pccdref}
E_{N} - E_{N+1} &= \langle 0 |  \hat{H}^{(\rm pCCD)}  | 0 \rangle - \langle 0 | a_a \hat{H}^{(\rm pCCD)} a_a^\dagger| 0 \rangle \notag\\
&= -\langle 0 | a_a^\dagger \left[ \hat{H}^{(\rm pCCD)} _N , a_a^\dagger \right] | 0 \rangle \notag\\
&= -f_{aa} + \sum_{k} t^{a\bar{a}}_{k\bar{k} } \langle k\bar{k} |a\bar{a} \rangle.
\end{align}
Note that the total pCCD correlation energy is $E^{\rm corr} = \sum_{kc} \braket{k\bar{k}}{c\bar{c}} t_{k\bar{k}}^{c\bar{c}}$.
Thus, in contrast to the conventional Koopmans' theorem, the modified versions---related to $\hat{H}_N^{(\rm pCCD)}$--- subtract or add the correlation energy contribution of the orbital in question ($i$ or $a$).
The terms in eqs.~\eqref{eqn:ip-pccdref} and \eqref{eqn:ea-pccdref} can also be deduced from the diagrammatic formulations of the IP- and EA-EOM-pCCD equations restricted to the 1-hole and 1-particle sectors, respectively (see Figure~\ref{fig:diagrams}).
Specifically, they correspond to the diagonal part of the corresponding IP/EA-EOM Hamiltonian.

\begin{table*}[t]
\caption{\label{tbl:tableequ}Summary of Koopmans' and Modified Koopmans' Equations for IPs, EAs, DIPs, and DEAs.}
\resizebox{1\textwidth}{!}{
\begin{ruledtabular}
\begin{tabular}{lll}
Orbital Energy & Koopmans' & Modified Koopmans' \\ 
IP            &   ${f}_{ii}$    &  $-{f}_{ii} - \sum\limits_{c}  {t^{c\bar{c}}_{i\bar{i}}} \braket{i\bar{i}}{c\bar{c}}$ \\ 
EA            &   $-{f}_{aa}$   &  $ -{f}_{aa} + \sum\limits_{k}  {t^{a\bar{a}}_{k\bar{k}}} \braket{k\bar{k}}{a\bar{a}}$ \\ 
\\
DIP ($m_s=1$) & $f_{ii} + f_{jj} - \langle{ij||ij}\rangle$  &  $-{f}_{ii} - f_{jj} + V_{ijij} - \sum\limits_{c}  t^{c\bar{c}}_{i\bar{i}} \braket{i\bar{i}}{c\bar{c}} - \sum\limits_{c}  t^{c\bar{c}}_{j\bar{j}} \braket{j\bar{j}}{c\bar{c}}$ \\
DIP ($m_s=0$) & $f_{ii} + f_{\bar{j}\bar{j}} - \braket{i\bar{j}}{i\bar{j}}$ & $-{f}_{ii} - f_{\bar{j}\bar{j}} + V_{i{\bar{j}}i{\bar{j}}} - \sum\limits_{c}  t^{c\bar{c}}_{i\bar{i}} \braket{i\bar{i}}{c\bar{c}} - \sum\limits_{c}  t^{c\bar{c}}_{j\bar{j}} \braket{j\bar{j}}{c\bar{c}} + \sum\limits_c t^{c\bar{c}}_{i\bar{i}} \braket{i\bar{i}}{c\bar{c}} \delta_{ij}$ \\
\\
DEA ($m_s=1$) & $f_{aa} + f_{bb} + \langle{ab||ab}\rangle $   &  $-{f}_{aa} - f_{bb} - V_{abab} + \sum\limits_{k}  {t^{a\bar{a}}_{k\bar{k}}} \braket{k\bar{k}}{a\bar{a}} + \sum\limits_{k}  {t^{b\bar{b}}_{k\bar{k}}} \braket{k\bar{k}}{b\bar{b}}$ \\
DEA ($m_s=0$) & $f_{aa} + f_{\bar{b}\bar{b}} + \braket{a\bar{b}}{a\bar{b}}$    & $-{f}_{aa} - f_{\bar{b}\bar{b}} - V_{a\bar{b}a\bar{b}} + \sum\limits_{k}  {t^{a\bar{a}}_{k\bar{k}}} \braket{k\bar{k}}{a\bar{a}} + \sum\limits_{k}  {t^{b\bar{b}}_{k\bar{k}}} \braket{k\bar{k}}{b\bar{b}} - \sum\limits_k t^{a\bar{a}}_{k\bar{k}}\braket{k\bar{k}}{a\bar{a}}\delta_{ab}$ \\ 

\end{tabular}
\end{ruledtabular}
}
\end{table*}

In a similar manner, we can adjust the energy expression for DIP and DEA.
The energy difference between the doubly-ionized state, where one electron each has been removed from the orbital pair $(i,j)$, and the reference state can be approximated as
{\small
\begin{align}
\label{eqn:dip-pccdref-1}
E_{N-2} - E_N &= \langle 0 | a_i^\dagger a_j^\dagger \hat{H}^{(\rm pCCD)} a_j a_i | 0 \rangle - \langle 0 | \hat{H}^{(\rm pCCD)} | 0 \rangle \notag\\
&= \langle 0 | a_i^\dagger a_j^\dagger (\hat{H}^{(\rm pCCD)} - E_{\text{ref}}) a_j a_i | 0 \rangle - \langle 0 | (\hat{H}^{(\rm pCCD)} - E_{\text{ref}}) | 0 \rangle \notag\\
&= \langle 0 | a_i^\dagger a_j^\dagger \hat{H}^{(\rm pCCD)} _N  a_j a_i | 0 \rangle 
- \langle 0 | \hat{H}^{(\rm pCCD)} _N | 0 \rangle \notag\\
&= \langle 0 | a_i^\dagger a_j^\dagger \left[ \hat{H}^{(\rm pCCD)} _N , a_j a_i \right] | 0 \rangle \notag\\
&= -f_{ii} - f_{jj} + V_{ijij} - \sum_{c} t^{c\bar{c}}_{i\bar{i} } \langle i\bar{i} |c\bar{c} \rangle - \sum_{c} t^{c\bar{c}}_{j\bar{j} } \langle j\bar{j} |c\bar{c} \rangle \notag\\
&= \textrm {IP}(i) + \textrm {IP}(j) + V_{ijij},
\end{align}
}
which is equivalent to the sum of the IPs of orbitals $i$ and $j$ plus the electron repulsion term $V_{ijij}$.
The above equation approximates the double ionization process in the case of the removal of electrons with the same spins.
Thus, it allows us to model the DIPs of the high-spin component of triplet states ($m_s=1$).
The double-ionization process of the $(i,\bar{j})$ spin-pair, on the other hand, encodes DIPs of, amongst others, the doubly-ionized singlet states ($m_s=0$).
For the $(i,\bar{j})$ spin case, we have
{\small
\begin{align}
\label{eqn:dip-pccdref-0}
E_{N-2} - E_N &= \langle 0 | a_i^\dagger a_{\bar{j}}^\dagger \hat{H}^{(\rm pCCD)} a_{\bar{j}} a_i | 0 \rangle - \langle 0 | \hat{H}^{(\rm pCCD)} | 0 \rangle \notag\\
&= \langle 0 | a_i^\dagger a_{\bar{j}}^\dagger \left[ \hat{H}^{(\rm pCCD)} _N , a_{\bar{j}} a_i \right] | 0 \rangle \notag\\
&= -f_{ii} - f_{jj} + V_{ijij} - \sum_{c} t^{c\bar{c}}_{i\bar{i} } \langle i\bar{i} |c\bar{c} \rangle - \sum_{c} t^{c\bar{c}}_{j\bar{j} } \langle j\bar{j} |c\bar{c} \rangle \notag\\
&\phantom{=}+ \sum_c t^{c\bar{c}}_{i\bar{i}} \langle{i\bar{i}}|{c\bar{c}}\rangle \delta_{ij} \notag\\
&= \textrm{IP}(i) + \textrm{IP}(\bar{j}) + V_{i\bar{j}i\bar{j}} + \sum_c t^{c\bar{c}}_{i\bar{i}} \langle{i\bar{i}}|{c\bar{c}}\rangle \delta_{ij},
\end{align}
}
which differs in the missing exchange term in $V_{i\bar{j}i\bar{j}}$ and the subtraction of the doubly-counted pCCD correlation terms associated with orbital $i$.

The approximate DEA energies are obtained in a similar manner.
For the triplet block ($m_s=1$), we have
{\small
\begin{align}
\label{eqn:dea-pccdref-1}
E_{N} - E_{N+2} &= \langle 0 |  \hat{H}^{(\rm pCCD)}  | 0 \rangle - \langle 0 | a_a a_{{b}} \hat{H}^{(\rm pCCD)} a_{{b}}^\dagger a_a^\dagger| 0 \rangle \notag\\
&= -\langle 0 | a_a a_{{b}}\left[ \hat{H}^{(\rm pCCD)} _N , a_{{b}}^\dagger a_a^\dagger \right] | 0 \rangle \notag\\
&= -f_{aa}  -f_{bb} -V_{abab} + \sum_{k} t^{a\bar{a}}_{k\bar{k} } \langle k\bar{k} |a\bar{a} \rangle + \sum_{k} t^{b\bar{b}}_{k\bar{k} } \langle k\bar{k} |b\bar{b} \rangle \notag\\
&= \textrm {EA}(a) + \textrm {EA}(b) - V_{abab},
\end{align}
}
while doubly-attached $m_s=0$ states are determined from
{\small
\begin{align}
\label{eqn:dea-pccdref-0}
E_{N} - E_{N+2} &= \langle 0 |  \hat{H}^{(\rm pCCD)}  | 0 \rangle - \langle 0 | a_a a_{\bar{b}} \hat{H}^{(\rm pCCD)} a_{\bar{b}}^\dagger a_a^\dagger| 0 \rangle \notag\\
&=- \langle 0 | a_a a_{\bar{b}}\left[ \hat{H}^{(\rm pCCD)} _N , a_{\bar{b}}^\dagger a_a^\dagger \right] | 0 \rangle \notag\\
&= -f_{aa}  -f_{{\bar{b}}{\bar{b}}} -V_{a{\bar{b}}a{\bar{b}}} + \sum_{k} t^{a\bar{a}}_{k\bar{k} } \langle k\bar{k} |a\bar{a} \rangle + \sum_{k} t^{b\bar{b}}_{k\bar{k} } \langle k\bar{k} |b\bar{b} \rangle \notag\\
&- \sum_{k} t^{a\bar{a}}_{k\bar{k} } \langle k\bar{k} |a\bar{a} \rangle \delta_{ab} \notag\\
&= \textrm{EA}(a) + \textrm{EA}(b) - V_{a{\bar{b}}a{\bar{b}}}  - \sum_{k} t^{a\bar{a}}_{k\bar{k} } \langle k\bar{k} |a\bar{a} \rangle \delta_{ab}.
\end{align}
}
As discussed for the IP/EA case above, the terms in eqs.~\eqref{eqn:dip-pccdref-1}-\eqref{eqn:dea-pccdref-0} can also be deduced from the diagrammatic formulations of the DIP- and DEA-pCCD equations restricted to the 2-hole and 2-particle sectors, respectively (see Figure~\ref{fig:diagrams}).
Specifically, they correspond to the diagonal part of the corresponding DIP/DEA-EOM Hamiltonian.
Table~\ref{tbl:tableequ}, summarizes all studied (D)IP/(D)EA expressions based on Koopmans' and modified Koopmans' theorems.

\begin{figure*}
\centering
\includegraphics[scale=0.87]{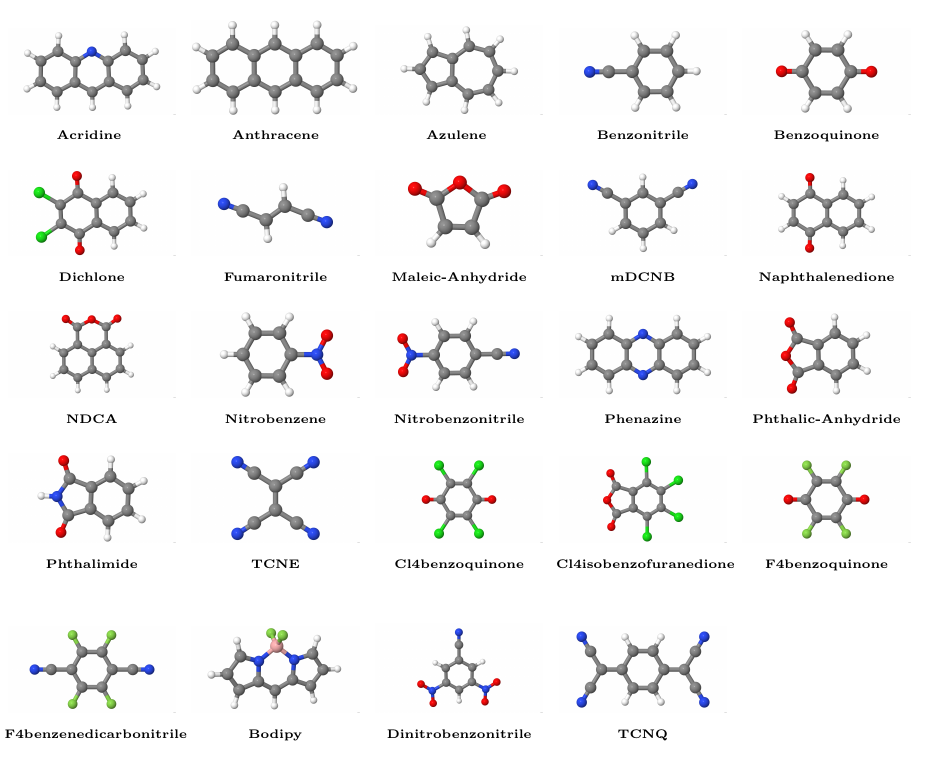}
\caption{An acceptor benchmark set composed of 24 organic molecules from Ref.~\citenum{knight2016accurate}.}
\label{fig:molecular-structures}
\end{figure*} 

\subsubsection{{Connection to the Extended Koopmans' theorem}}
{Within the EKT the approximate orbital energies are determined by solving the following secular equation} 
\begin{equation}
{\bf FC} -\epsilon {\bf \gamma C}  ={\bf 0}, 
\end{equation}
{in which $C$ is the matrix of eigenvectors, $\epsilon$ is the diagonal matrix of eigenvalues (orbital energies), $\gamma_{pq}$ is the one-particle reduced density matrix (1-RDM), }
\begin{equation}\label{eq:1-rdm}
\gamma_{pq}=\langle \Psi^N|\hat{a}_p^{\dagger}\hat{a}_q|\Psi^N \rangle, 
\end{equation}
{and ${\bf F}$ is the so-called generalized Fock matrix (aka Lagrangian), }
\begin{equation}
\text{F}_{pq}=-\langle \Psi^N|\hat{a}_p^{\dagger}[\hat{H},\hat{a}_q]|\Psi^N\rangle. 
\end{equation}
 

\subsubsection{EOM-pCCD type methods}
We utilized the equation of motion (EOM) formalism~\cite{rowe-eom, bartlett-eom} on top of the pCCD reference wave function to generate reference data for the pCCD-based IPs, EAs, DIPs, and DEAs.~\cite{ip-pccd, galyńska2024benchmarking} 
Within the (D)IP/(D)EA-EOM-pCCD formalism, we remove or attach electrons to the closed-shell pCCD reference function through a linear ansatz to parametrize the $k$-th ionized, doubly ionized, attached, and doubly attached state
\begin{equation}\label{eq:R-operator}
\ket{\Psi_k} = \hat R(k)\ket{\textrm{pCCD}}
\end{equation}
{where the operator ${\hat{R}(\mathnormal{k})}$ generates the targeted state $\mathnormal{k}$ from the initial pCCD reference state. The explicit form of the operator ${\hat{R}(\mathnormal{k})}$ determines the EOM flavor. Open-shell species are, for instance, accessible by restricting ${\hat{R}(\mathnormal{k})}$ to target either ionized or electron-attached states with respect to the pCCD reference wave function. To simplify the notation, we will henceforth omit the $\mathnormal{k}$ dependence of the ${\hat{R}}$ operator. The operator $\hat{R}$ is typically decomposed into different parts, where each part is characterized by the number of particle (electron creation) and hole (electron annihilation) operators it contains. To target quantum states with a single unpaired electron, the single IP-EOM  formalism\mbox{~\cite{musial2003equation}} can be employed, in which the $\hat{R}$ operator is composed of 1 hole (h) operator, 2 hole and 1 particle (p) operators, and so on.,}
\begin{equation}
    \label{eqn:R-ip-pccd} 
    \begin{aligned}
        \hat{R}^{\rm IP} &= \sum_{i} r_i \hat{a}_i + \frac{1}{2} \sum_{i j a} r^a_{ij} \hat{a}_a^\dagger \hat{a}_j \hat{a}_i + \dots \\ 
        &= \hat{R}_{1h} + \hat{R}_{2h1p} + \dots,
    \end{aligned}
\end{equation}
{where $r_i$ describes the linear excitation operator. Likewise, quantum states with two or more unpaired electrons can be targeted using the double (D)IP-EOM model, where $\hat{R}$ operator is composed of 2h, 3h1p, etc., }
\begin{equation}
      \label{eqn:equation21}
  \begin{aligned}
 \hat R^{\rm DIP} = \frac{1}{2} \sum_{i}r_{ij}{\hat j\hat i}+ \frac{1}{6}\sum_{ijk{a}}{r}^a_{ijk} \hat{a}^\dag \hat k \hat j \hat{i}+ \dots ~ = \hat R_{2h}+\hat R_{3h1p}+\dots ~ 
 \end{aligned}
\end{equation}
{In the EA-EOM (Electron Attachment Equation-of-Motion) formalism \mbox{\cite{nooijen1995equation, deaeomccsdt}}, the linear excitation operator, ${\hat{R}}$, has the form}
\begin{equation}
      \label{eqn:equation24}
  \begin{aligned}
 \hat R^{\rm EA} = \sum_{a}r^a\hat{a}_a^\dag+ \frac{1}{2}\sum_{ab{j}}{r}^{ab}_j\hat{a}_a^\dag \hat{a}_b^\dag \hat a_j+ \dots ~ = \hat R_{1p}+\hat R_{2p1h}+\dots ~ 
 \end{aligned}
\end{equation} 
{The Double Electron Attachment DEA-EOM-CC formalism\mbox{\cite{gulania2021, ip-pccd}} extends the approach to target doubly-electron-attached states by incorporating additional particle terms in the ${\hat{R}}$ operator.}
\begin{equation}
      \label{eqn:equation25}
  \begin{aligned}
 \hat R^{\rm DEA} = \sum_{ab}r^{ab}\hat{a}_a^\dag \hat{a}_b^\dag+ \frac{1}{6}\sum_{abc{k}}{r}^{abc}_k \hat{a}_a^\dag \hat{a}_b^\dag \hat{a}_c^\dag \hat{a}_j + \dots ~ = \hat R_{2p}+\hat R_{3p1h}+\dots ~ 
 \end{aligned}
\end{equation}
{The ionized and attached states are generated by solving the associated EOM equations}
\begin{equation}
      \label{eqn:equation22}
  [\hat{H}_N,\hat R ] \ket{\textrm{pCCD}} =  \omega \hat R \ket{\textrm{pCCD}}
\end{equation}
{where ${\omega = \Delta{E}- \Delta{E_0} }$ is the change in energy associated with the single/double ionization energy (electron attachment) process relative to the pCCD ground-state energy, and ${\hat{H}_N = \hat{H} - {\langle \Phi_0|\hat{H}| \Phi_0\rangle}}$ is the normal-product form of the Hamiltonian. The above equation can be written as }
\begin{equation}
      \label{eqn:equation23}
 {\cal H}^{\rm{pCCD}}_N \hat R\ket{\Phi_0} =  \omega \hat R \ket{\Phi_0}.
\end{equation}
{The ionization energies and electron attachments are obtained as the eigenvalues of a non-Hermitian matrix, which can be iteratively diagonalized to determine the lowest-lying ionized and attached states. This is based on the introduction of the similarity transformed Hamiltonian of the pCCD model in its normal-product form $  {\cal H}^{\rm{pCCD}}_N= e^{- \hat{T}^{\rm{pCCD}}} \hat{H}_N e^{\hat{T}^{\rm{pCCD}}}$.}
\begin{figure*}
\centering
\includegraphics[width=1.0\textwidth]{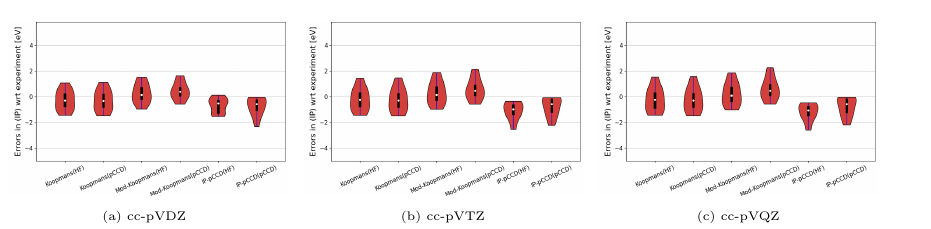}
\caption{Errors in IPs for all investigated Koopmans' flavors and the IP-EOM-pCCD(2h,1p) (denoted as IP-pCCD) model for various basis set sizes wrt experimental data {for selected closed-shell atoms.}}
\label{fig:errors-ip-atoms}
\end{figure*} 
\section{Computational details}\label{sec:comput-det}
All implementations and calculations have been performed in a developer version of the PyBEST v2.1.0dev0 software package.~\cite{pybest-paper, boguslawski2024pybest}
We benchmarked IPs, EAs, DIPs, and DEAs for eight neutral atoms (He, Be, Ne, Mg, Ca, Ar, Kr, and Zn) against the (D)IP/(D)EA/-EOM-pCCD data.~\cite{ip-pccd, galyńska2024benchmarking, galyńska2024exploring}
We used the correlation consistent basis set series, including cc-pVDZ, cc-pVTZ, and cc-pVQZ~\cite{dunning1989gaussian} and two sets of molecular orbitals: (i) canonical Hartree-Fock orbitals (abbreviated as HF) and (ii) variationally optimized natural pCCD orbitals (abbreviated as pCCD~\cite{boguslawski2014efficient}).
We also investigated a data set of 24 organic acceptors for which experimental references exist.~\cite{knight2016accurate}
The optimized xyz structures are taken from Ref.~\citenum{knight2016accurate}.
These molecules are collected in Figure~\ref{fig:molecular-structures}.
For the molecular benchmark data set, we used the cc-pVDZ~\cite{dunning1989gaussian}, {aug-cc-pVDZ, and aug-cc-pVTZ}~\cite{aug-cc-pvtz} basis sets and benchmarked canonical HF and natural pCCD orbitals.

In all the IP-EOM-pCCD calculations, we used the 2h1p and 1h in the $\hat{R}$ operator (cf. eq.~\ref{eq:R-operator}) and for the EA-EOM-pCCD model, only the 2p1h model. 
For the DIP/DEA-EOM-pCCD, we considered 3h1p and 3p1h in the $\hat{R}$ operator, respectively. 
In all cases, we computed 3-6 lowest-lying roots and used a frozen core approximation (1s for Be, C, N, O, and F; 1s--2p for Ca, Si, P, S, Cl, and Ar; 1s--3p  for Zn and Kr). 
Due to convergence issues, in some cases, we were forced to drop the core approximation. 
Such cases are explicitly noted in the tables.  
Our preliminary tests suggest that freezing core orbitals has a negligible effect on the computed properties.

\section{Results and discussion}\label{sec:results}
In the following, we will investigate the performance of our new computational models for predicting orbital and pair orbital energies of selected atoms and a benchmark set of prototypical organic donors~\cite{knight2016accurate} illustrated in Figure~\ref{fig:molecular-structures}.
First, we set our focus on single orbital energies and their ability to approximate IP and EA. 
Later, we investigate the performance of HF- and pCCD-based pair orbital energies for atoms and molecules.
\begin{table*}[ht!]
\caption{
Comparison of experimental IPs (in eV) of selected closed-shell atoms with different theoretical models and basis set sizes. 
HF and pCCD indicate canonical Hartree--Fock and orbital-optimized natural pCCD orbitals, respectively. 
ncore denotes the number of frozen core orbitals when it differs from a default value.
}
\label{tbl:table-ip-atom}
\resizebox{1\textwidth}{!}{
\begin{ruledtabular}
\begin{tabular}{llccccccc}
\Lower{Atom} & \Lower{Basis set} & \Lower{Exp.~\cite{nist}} & \multicolumn{2}{c}{Koopmans} & \multicolumn{2}{c}{Modified Koopmans} & \multicolumn{2}{c}{IP-EOM-pCCD (2h, 1p)} \\
\cline{4-9}
& & & HF & pCCD & HF & pCCD & HF & pCCD \\

He & cc-pVDZ &  & 24.88 & 24.89 & 25.76 & 25.77 & 24.32 & 24.33 \\
   & cc-pVTZ &  & 24.97 & 24.97 & 25.75 & 26.03 & 24.24 & 24.53 \\
   & cc-pVQZ & 24.59 & 24.98 & 24.97 & 25.66 & 26.08 & 24.13 & 24.56 \\
&\\

Be (ncore=0) & cc-pVDZ &  & 8.41 & 8.34 & 9.17 & 9.56 & 8.84 & 9.29 \\
             & cc-pVTZ &  & 8.42 & 8.33 & 9.05 & 9.57 & 8.67 & 9.26 \\
             & cc-pVQZ & 9.32 & 8.42 & 8.34 & 8.98 & 9.58 & 8.59 & 9.25 \\
&\\

Ne & cc-pVDZ &  & 22.64 & 22.67 & 23.09 & 23.18 & 21.68$^a$ & 19.25 \\
   & cc-pVTZ &  & 23.01 & 23.03 & 23.45 & 23.69 & 19.02$^a$ & 19.33 \\
   & cc-pVQZ & 21.56 & 23.10 & 23.13 & 23.42 & 23.83 & 18.96$^a$ & 19.36 \\
&\\   

Mg (ncore=1) & cc-pVDZ &  & 6.88 & 6.83 & 7.43 & 7.73 & 7.43 & 7.51 \\
             & cc-pVTZ &  & 6.89 & 6.83 & 7.34 & 7.75 & 7.04 & 7.50 \\
             & cc-pVQZ & 7.65 & 6.89 & 6.83 & 7.25 & 7.75 & 6.94 & 7.50 \\
&\\

Ar & cc-pVDZ &  & 16.00 & 15.96 & 16.34 & 16.37 & 14.35$^a$ & 14.40 \\
   & cc-pVTZ &  & 16.06 & 16.00 & 16.44 & 16.54 & 14.16$^a$ & 14.29 \\
   & cc-pVQZ & 15.76 & 16.08 & 16.02 & 16.41 & 16.58 & 14.05 & 14.27 \\
&\\

Ca (ncore=5) & cc-pVDZ &  & 5.32 & 5.28 & 5.78 & 6.03 & 5.57 & 5.86 \\
             & cc-pVTZ &  & 5.32 & 5.28 & 5.74 & 6.05 & 5.51 & 5.86 \\
             & cc-pVQZ & 6.11 & 5.12 & 5.28 & 5.64 & 6.05 & 5.38 & 5.86 \\
&\\

Zn & cc-pVDZ &  & 7.96 & 7.92 & 8.44 & 8.82 & 8.07 & 8.49 \\
   & cc-pVTZ &  & 7.96 & 7.92 & 8.43 & 8.82 & 8.05 & 8.48 \\
   & cc-pVQZ & 9.39 & 7.96 & 7.92 & 8.38 & 8.83 & 7.99 & 8.48 \\
&\\

Kr & cc-pVDZ &  & 14.17 & 14.15 & 14.45 & 14.51 & 12.47$^a$ & 12.95 \\
   & cc-pVTZ &  & 14.25 & 14.22 & 14.56 & 14.66 & 12.65 & 12.81 \\
   & cc-pVQZ & 14.00 & 14.26 & 14.22 & 14.52 & 14.70 & 12.56 & 12.80 \\
   &\\
\end{tabular}
\end{ruledtabular}
\footnotetext[1]{The IP-EOM-pCCD numbers indicate convergence issues with the Davidson algorithm and are converged only up to \(1 \times 10^{-3}\) $E_h$.}
}
\end{table*}


\begin{table*}[ht!]
\caption{\label{tbl:table-ea-atom} Comparison of theoretical EAs (in eV) in different basis sets for selected closed-shell atoms. 
HF and pCCD indicate canonical Hartree--Fock and orbital-optimized natural pCCD orbitals, respectively. 
ncore denotes the number of frozen core orbitals when it differs from a default value.}
\resizebox{1.01\textwidth}{!}{
\begin{ruledtabular}
\begin{tabular}{llllllll}

\Lower{Atom} & \Lower{Basis set} & \multicolumn{2}{c}{Koopmans} & \multicolumn{2}{c}{Modified Koopmans} & \multicolumn{2}{c}{EA-EOM-pCCD (2p, 1h)} \\ 
\cline{3-8}

& & HF & pCCD & HF & pCCD & HF & pCCD \\ 

He
& cc-pVDZ & 38.03 & 38.02 & 38.42 & 38.42 & 37.36 & 37.35 \\
& cc-pVTZ & 17.32 & 43.85 & 17.46 & 44.26 & 17.04 & 17.12 \\
& cc-pVQZ & 13.51 & 42.90 & 13.59 & 43.31 & 13.23 & 13.32 \\
&\\

Be (ncore=0) 
& cc-pVDZ & 1.59 & 3.20 & 1.77 & 3.58 & 1.06 & 1.23 \\
& cc-pVTZ & 1.36 & 3.20 & 1.48 & 3.58 & 0.69 & 0.91 \\
& cc-pVQZ & 1.22 & 3.25 & 1.32 & 3.63 & 0.54 & 0.79 \\
&\\

Ne 
& cc-pVDZ & 46.11 & 46.09 & 46.35 & 46.37 & 44.19$^a$ & 44.25 \\
& cc-pVTZ & 29.90 & 51.44 & 30.06 & 51.74 & 28.50$^a$ & 28.60 \\
& cc-pVQZ & 22.01 & 51.43 & 22.06 & 51.74 & 20.91$^a$ & 21.00 \\
&\\

Mg (ncore=1)
 & cc-pVDZ & 1.22 & 3.06 & 1.32 & 3.33 & 0.84 & 0.96 \\
 & cc-pVTZ & 1.00 & 3.00 & 1.06 & 3.27 & 0.51 & 0.66 \\
 & cc-pVQZ & 0.79 & 3.00 & 0.82 & 3.27 & 0.32 & 0.48 \\
 &\\

Ar
 & cc-pVDZ & 21.69 & 21.71 & 21.75 & 21.81 & 20.55$^a$ & 20.61 \\
 & cc-pVTZ & 14.97 & 22.11 & 15.00 & 22.37 & 13.97$^a$ & 17.01 \\
 & cc-pVQZ & 10.53 & 21.23 & 10.55 & 21.50 & 9.72$^a$ & 11.23 \\
 &\\

Ca (ncore=5)
& cc-pVDZ & 0.67 & 1.88 & 0.76 & 2.11 & 0.10 & 0.23 \\
& cc-pVTZ & 0.62 & 1.77 & 0.70 & 2.00 & -0.02 & 0.13 \\
& cc-pVQZ & -0.44 & 1.57 & -0.45 & 1.57 & -4.08 & 0.01 \\
&\\

Zn
 & cc-pVDZ & 1.49 & 3.86 & 1.57 & 4.14 & 1.03 & 1.16 \\
 & cc-pVTZ & 1.47 & 3.93 & 1.55 & 4.21 & 0.98 & 1.13 \\
 & cc-pVQZ & 1.28 & 3.92 & 1.34 & 4.21 & 0.71 & 0.87 \\
 &\\

Kr
 & cc-pVDZ & 19.70 & 19.72 & 19.74 & 19.77 & 17.98$^a$ & 17.99 \\
 & cc-pVTZ & 11.39 & 18.42 & 11.41 & 18.64 & 10.54$^a$ & 11.07 \\
 & cc-pVQZ & 7.22 & 17.72 & 7.23 & 17.94 & 6.61$^a$ & 8.28 \\
 &\\


\end{tabular}
\end{ruledtabular}
\footnotetext[1]{Convergence issues with the Davidson algorithm led to imaginary eigenvalues, resulting in calculations converged only to \(1 \times 10^{-3}\) $E_h$.}
}
\end{table*}

\subsection{Orbital Energies}
\subsubsection{Atoms}
\balance 
We start our discussion by investigating IPs and EAs of the He, Be, Ne, Mg, Ar, Ca, Zn, and Kr atoms.
A reliable experimental IP reference~\cite{nist} exists for these systems. 
They all have closed-shell electronic structures which are dominated by different electron correlation effects. 
While the He, Ne, Ar, and Kr noble gases are prime examples of weakly correlated systems, Be, Mg, Ca, and Zn exhibit somehow mixed degrees of weak and strong electron correlation effects. 
To that end, these atoms provide a good testing ground for new electronic structure methods. 

Table~\ref{tbl:table-ip-atom} collects the experimental and calculated IPs using various theoretical models: Koopmans' theorem, modified Koopmans' theorem, and IP-EOM-pCCD. 
All these methods used two sets of orbitals: canonical Hartree--Fock and natural pCCD orbitals. 
In general, the pCCD orbitals are localized in nature, but they remain almost identical to the canonical HF orbitals in isolated atoms.
The computed IPs generally agree well with experimental values, and the agreement is better for larger basis sets. 
The only exception is the IP-EOM-pCCD(2h,1p) approach, which struggles for heavier noble gases (cf. IPs for Ne, Ar, and Kr in Table~\ref{tbl:table-ip-atom}), where we experienced numerical difficulties in the Davidson diagonalization with degenerate orbitals. 

For an in-depth analysis of the performance of individual methods in various basis sets, we refer to the violin plots depicted in Figure~\ref{fig:errors-ip-atoms}.
Specifically, these plots illustrate the locality, spread, skewness, and overall distribution of errors w.r.t. experimental reference.  
Figure~\ref{fig:errors-ip-atoms} suggests that IPs from the modified Koopmans' theorem provide the most reliable results across different basis sets. 
The modified Koopmans' errors are mostly centered around the reference values; the spread is small and decreases with basis set size.
The modified Koopmans(pCCD) approach is the most accurate model and outperforms its modified Koopmans(HF) variant. 
The Koopmans' approach exhibits slightly larger errors and broader distributions. 
IP-EOM-pCCD(2h1p) provides overall smaller spreads but systematically underestimates IPs. 
A statistical analysis of the computed IPs, including mean absolute errors, standard deviation, mean error, mean absolute error, root mean square error, and mean percentage errors, are summarized in Table S1 of the SI.

Table~\ref{tbl:table-ea-atom} collects electron affinities of selected closed-shell atoms calculated using the same methods and basis sets as for IPs. 
Among the atoms studied, He and Ne show the largest EA values, ranging from approximately 37-46 eV for He and 44-51 eV for Ne across all basis sets. 
In contrast, Be, Mg, and Ca exhibit considerably lower EAs, typically below 4 eV. 
Here, we observe a large basis set dependence compared to IPs.
That dependence is less pronounced for the pCCD-optimized orbitals. 
The pCCD-optimized orbitals consistently provide higher EA values compared to HF, particularly for heavier atoms. 
The dependence on orbitals is somehow reduced in the EA-EOM-pCCD approach. 
In general, EAs are getting smaller in larger basis sets. 
One exception is the Ca atom, which becomes unstable in a larger basis set.  
Since there are no reliable experimental reference values for EAs, we will focus on comparing our models to the EA-EOM-pCCD(pCCD) approach. 
We should stress here that recent benchmark studies on EA-EOM-pCCD revealed an average error of up to 1 eV for EA's.~\cite{galyńska2024benchmarking}
Even though our reference might not be perfect, this is a good theoretical reference for our simple pCCD-based models. 
The statistical analysis presented in Table S2 of the SI suggests that Koopmans(HF) and modified Koopmans(HF) are the closest to the EA-EOM-pCCD reference. 

\begin{figure*}[t]
\centering
\includegraphics[width=1.0\textwidth]{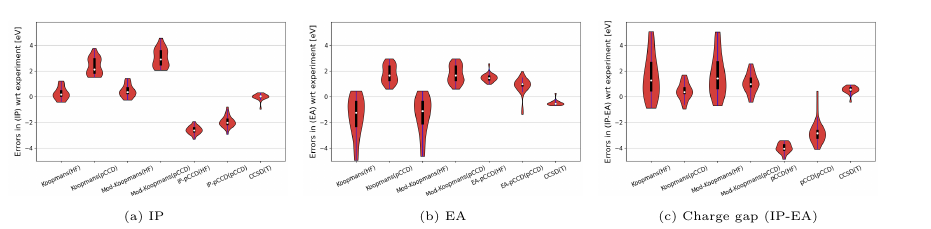}
\caption{Errors in IPs, EAs, and charge gaps (in eV) w.r.t. experiment using aug-cc-pVDZ basis set.
Note that due to the lack of some experimental data IPs are calculated for 21 molecules, EAs for 22, and charge gap for 20 molecules (see also Tables~S5, S6 and, {S13} of the SI).
}
\label{fig:errors-ip-molecules}
\end{figure*} 
\subsubsection{Molecules}
Accurate predictions of IPs, EAs, and optical gaps ($\Delta c = \text{IP} - \text{EA}$) are essential for molecules that act as fundamental components in organic electronics. 
The performance of our methods for IPs, EAs, and charge gaps for acceptor molecules (cf. Figure~\ref{fig:molecular-structures}) is summarized in Figure~\ref{fig:errors-ip-molecules}. 
Tables S5 and S6 of the SI collect all the individual IPs and EAs, respectively. 
It is evident from Figure~\ref{fig:errors-ip-molecules}a that the most reliable IPs are coming from CCSD(T). 
Yet, the Koopmans(HF) and modified Koopmans(HF) values are also very good; the errors are centered around the references and usually do not exceed 0.5 eV. 
Applying the Koopmans and modified Koopmans theorem on top of pCCD orbitals worsens the results for IPs. 
The spread is larger and IPs are overestimated by a few eV.  
The picture is reversed for the EAs presented in Figure~\ref{fig:errors-ip-molecules}b, where the pCCD orbitals are more reliable than the canonical HF ones. 
In that case, the performance of Koopmans(pCCD) and modified Koopmans(pCCD) is similar to that of IPs. 
Together, these methods lead to quite satisfactory charge gaps as depicted in Figure~\ref{fig:errors-ip-molecules}. 
Working with pCCD orbitals provides a more balanced description of the occupied and virtual orbitals (due to the inclusion of electron correlation effects) than in the canonical HF case.
The virtual orbitals in HF theory do not have any physical meaning, and as a consequence, their description is usually inferior. 
This could also be noticed in the CCSD(T) results, where its performance for EAs is slightly worse than that of IPs. 
We should stress here that although Koopmans(pCCD) and modified Koopmans(pCCD) charge gaps provide larger errors and broader spreads w.r.t. experiment than CCSD(T), their computational cost is much smaller (o$^2$v$^2$ vs. o$^2$v$^4$).
{For example, the ME w.r.t. experiment of pCCD-based charge gaps is 0.42 eV, and slightly better than the CCSD(T) ME of 0.54 eV using the aug-cc-pVDZ basis set.
However, pCCD results in a larger standard deviation (0.63 eV) compared to CCSD(T) (0.30 eV).
The results deteriorate for aug-cc-pVTZ basis set, but are still better then for cc-pVDZ (see Table S14 of the SI). }

\begin{figure}[t]
\centering
\includegraphics[width=0.45\textwidth]{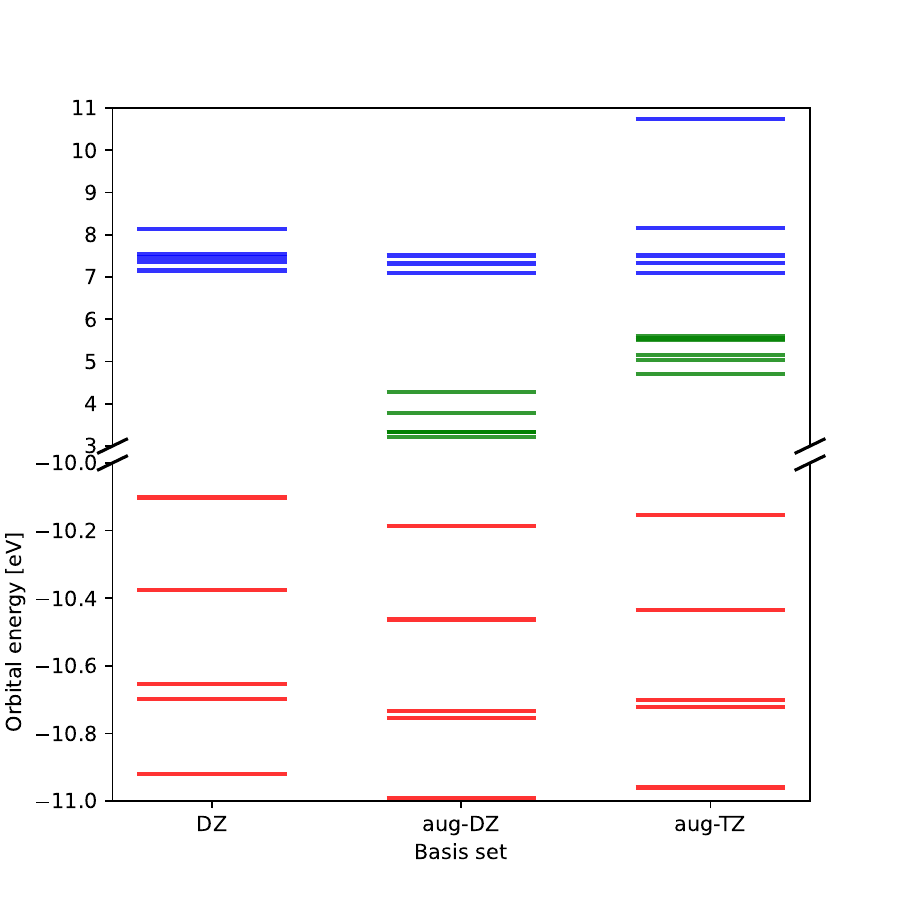}
\caption{{Evolution of Koopmans(pCCD) valence orbital energies for azulene molecule across different basis set sizes. The red lines correspond to the occupied, blue, and green virtual orbitals. The local virtual orbitals are denoted blue, and green colors denote the Rydberg orbitals.} 
}\label{fig:rydberg-orbital-energies}
\end{figure}

{While the inclusion of augmented functions does not effect much the IPs, it improves the EAs as the attached electrons occupy the low-lying Rydberg-type orbitals first. 
Figure\mbox{~\ref{fig:rydberg-orbital-energies}} provides an illustrative example of such behavior. }

{The overestimation of IPs by the IP-EOM-pCCD(pCCD) model is well-documented in the literature\mbox{~\cite{ip-pccd, galyńska2024benchmarking}} and attributed to the unbalanced treatment of electron correlation effects in the reference oo-pCCD state and the ionized state, which exhibits greater correlation via the $R$ operator\mbox{~\eqref{eqn:R-ip-pccd}}  .
A similar observation, but to a lesser extent, has been observed for EA-EOM-pCCD.\mbox{~\cite{galyńska2024exploring}} Based on this considerations, the overestimation of IPs and EAs by Koopmans(pCCD) and modified Koopmans(pCCD) is expected.} 

Additional information on statistical data, including standard deviation, mean error, mean absolute error, root mean square error, and mean percentage errors, are listed in Tables S9, S10, S11, S12, and {S14} of the SI.

\subsection{Pair orbital energies}
\subsubsection{Atoms}
\begin{figure*}[ht!]
\centering
 \includegraphics[width=1.0\textwidth]{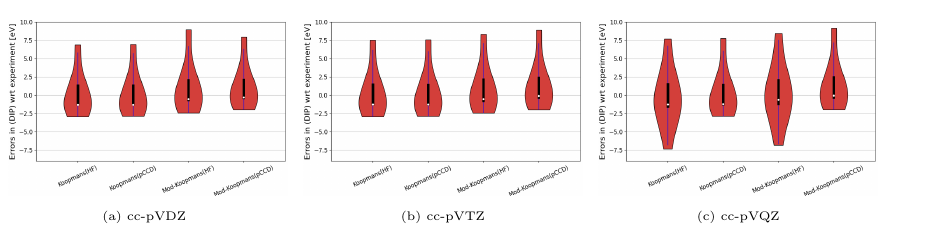}
\caption{Errors in DIPs for all investigated Koopmans' flavors for various basis set sizes {w.r.t. experimental data for selected closed-shell atoms. Note that due to convergence issues with the Davidson algorithm for some systems, DIP-EOM-pCCD results are not included.}
}
\label{fig:errors-dip-atoms}
\end{figure*}

Table~\ref{tbl:table-dip-atom} summarizes the lowest-lying experimental and calculated DIPs for selected closed-shell atoms using various theoretical models, three basis sets, and two sets of orbitals. 
As shown in Table~\ref{tbl:table-dip-atom}, we face numerical difficulties in convergence for Davidson diagonalization in the DIP-EOM-pCCD approach.
Most of these convergence issues are leveraged by using pCCD-optimized natural orbitals.
Despite these issues, the computed DIPs agree qualitatively with experimental values.
The performance of individual Koopmans' methods across different basis sets is summarized in Figure~\ref{fig:errors-dip-atoms}.
Most of the data is centered around the reference experimental values, but the spread is large.
Such a large spread originates mainly from the Ne results, for which our methods differ by up to 7 eV.
Generally, the computed DIPs do not depend on the basis set size. 
The only exceptions are the Koopmans(HF) and modified Koopmans(HF) methods for Ca using a cc-pVQZ basis (cf. Table~\ref{tbl:table-dip-atom}). 
Additional statistical errors of calculated DIPs, including standard deviation, mean error, mean absolute error, root mean square error, and mean percentage error, are provided in Tables S3 of the SI.

Table~\ref{tbl:table-dea-atom} collects DEAs, utilizing the same basis sets and methods employed for DIPs focusing on the lowest-lying singlet and triplet electronic configurations. 
The only exception is the Zn atom, where we used the $^1S_0$ state instead of the lowest-lying triplet state, for which we encountered convergence difficulties. 
The results show that both the Koopmans(HF) and modified Koopmans(HF) yield significantly lower differences w.r.t. DEA-EOM-pCCD(pCCD) reference values.
Employing natural pCCD orbitals leads to a dramatic increase in DEAs towards larger basis sets.  
While DEAs for Koopmans(HF) and modified Koopmans(HF) follow the DEA-EOM-pCCD decreasing trend for larger basis set, the opposite is true for  Koopmans(pCCD) and modified Koopmans(pCCD). 
That substantial difference is largely attributed to hybridization effects and symmetry breaking within the pCCD orbital optimization.
Without experimental data for validation, it remains challenging to determine which approach is superior.
The statistical analysis for DEAs, including standard deviation, mean errors, mean absolute errors, root mean square errors, and mean percentage errors for all methods evaluated in Table S4 of SI.
\begin{table*}[ht!]
\caption{\label{tbl:table-dip-atom}
Comparison of experimental DIPs (in eV) of selected closed-shell atoms with different theoretical models and basis set sizes. 
HF and pCCD inidicate canonical Hartree--Fock and orbital-optimized natural pCCD orbitals, respectively. 
ncore denotes the number of frozen core orbitals when it differs from a default value.
}
    \resizebox{1.01\textwidth}{!}{
    \begin{ruledtabular}
    \begin{tabular}{llllllllllll}

\Lower{Atom}  & \Lower{Basis set} & \Lower{Terms} & \Lower{i}& \Lower{j}& \Lower{Exp.~\cite{nist}}& \multicolumn{2}{c}{Koopmans} & \multicolumn{2}{c}{Modified Koopmans} & \multicolumn{2}{c}{DIP-EOM-pCCD(3h,1p)}  \\ \cline{7-12}
 & & & & &   & HF & pCCD & HF & pCCD & HF & pCCD   \\ 

He
  & cc-pVDZ  &  $\ce{^{1}S_{0}}$  & ${1s_i}$  & ${1s_{\bar{j}}}$ &  & 77.69 & 77.69 & 78.58 & 78.58 & $^a$-- & $^a$-- \\
  & cc-pVTZ  &  &  &  & & 77.86 & 77.86 & 78.64 & 78.92 & $^a$-- & $^a$-- \\
  & cc-pVQZ  &  &  &  & 79.01 & 77.87 & 77.87 & 78.55 & 78.98 & $^a$-- & $^a$-- \\
 
 &\\

Be (ncore=0)
  & cc-pVDZ  &  $\ce{^{1}S_{0}}$  & ${2s_i}$  & ${2s_{\bar{j}}}$ &  & 26.17 & 26.15 & 26.93 & 27.42 & 26.93 & 27.38 \\
   & cc-pVTZ  &  &  &  & & 26.17 & 26.19 & 26.81 & 27.43 & 26.78 & 27.38 \\
  & cc-pVQZ  &  &  &  & 27.53 & 26.17 & 26.21 & 26.73 & 27.45 & 26.69 & 27.35 \\

 &\\

Ne
  & cc-pVDZ  & $\ce{^{3}P_{2}}$  &  ${2p_z}$ & ${2p_y}$  & & 69.39 & 69.45 & 71.48 & 70.47 & $^a$-- & $^a$-- \\
   & cc-pVTZ  &  &  &  & & 70.06 & 70.12 & 70.86 & 71.43 & $^a$-- & 59.21 \\
  & cc-pVQZ  &  &  &  & 62.53 & 70.23 & 70.29 & 70.99 & 71.69 & 58.53$^a$ & 59.32 \\

 &\\

Mg (ncore=1)
  & cc-pVDZ  &  $\ce{^{1}S_{0}}$  & ${3s_i}$   & ${3s_{\bar{j}}}$ & & 21.36 & 21.35 & 21.90 & 22.25 & 21.86 & 22.22 \\
  & cc-pVTZ  &  &  &  & & 21.36 & 21.35 & 21.81 & 22.27 & 21.75 & 22.21 \\
  & cc-pVQZ  &  &  &  & 22.68 & 21.36 & 21.35 & 21.72 & 22.27 & 21.65$^a$ & 22.20 \\

 &\\

Ar
  & cc-pVDZ  &  $\ce{^{3}P_{2}}$  &  ${3p_z}$ &  ${3p_y}$ & & 45.42 & 45.35 & 46.21 & 46.19 & $^a$-- & 40.88 \\
   & cc-pVTZ  &  &  &  & & 45.52 & 45.42 & 46.28 & 46.54 & $^a$-- & 40.88 \\
  & cc-pVQZ  &  &  &  & 43.39 & 45.55 & 45.45 & 46.26 & 46.63 & $^a$-- & 40.92 \\

 &\\

Ca (ncore=5)
  & cc-pVDZ  &  $\ce{^{1}S_{0}}$  & ${4s_i}$  & ${4s_{\bar{j}}}$ & & 16.47 & 16.47 & 16.94 & 17.22 & 16.88 & 17.17 \\
  & cc-pVTZ  &  &  &  & & 16.47 & 16.46 & 16.90 & 17.23 & 16.80 & 17.15 \\
  & cc-pVQZ  &  &  &  & 17.98 & 10.45 & 16.45 & 11.14 & 17.23 & 16.68 & 17.15 \\

 &\\

Zn
  & cc-pVDZ  &  $\ce{^{1}S_{0}}$ & ${4s_i}$  & ${4s_{\bar{j}}}$ & & 24.43 & 24.48 & 24.92 & 25.38 & 24.62$^a$ & 25.04\\
   & cc-pVTZ  &  &  &  & & 24.43 & 24.48 & 24.90 & 25.38 & 24.60 & 25.04 \\
  & cc-pVQZ  &  &  &  & 27.36 & 24.44 & 24.48 & 24.86 & 25.39 & 24.54$^a$ & 25.04 \\

 &\\

Kr
  & cc-pVDZ  &  $\ce{^{3}P_{2}}$  &  ${4p_z}$  & ${4p_y}$ & & 39.70 & 39.67 & 40.35 & 40.41 & 36.19$^a$ & 36.30 \\
   & cc-pVTZ  &  &  &  & & 39.84 & 39.78 & 40.45 & 40.70 & 35.87 & 36.18 \\
  & cc-pVQZ  &  &  &  & 38.36 & 39.85 & 39.80 & 40.37 & 40.74 & $^a$-- & 36.17 \\

 &\\

\end{tabular}
\end{ruledtabular}

\footnotetext[1]{The DIP calculations for He, Ne, Ar, and Kr did not converge due to numerical difficulties. Convergence issues with the Davidson algorithm led to imaginary eigenvalues, resulting in calculations converged only to \(1 \times 10^{-3}\).}
}
\end{table*}


\begin{table*}[ht!]
\caption{\label{tbl:table-dea-atom} Comparison of theoretical DEAs (in eV) in different basis sets for selected closed shell atoms. 
HF and pCCD indicate canonical Hartree--Fock and orbital-optimized natural pCCD orbitals, respectively. 
ncore denotes the number of frozen core orbitals when it differs from a default value.}
    \resizebox{1.01\textwidth}{!}{
    \begin{ruledtabular}
    \begin{tabular}{lllllllllll}

\Lower{Atom}  & \Lower{Basis set} & \Lower{Terms}& \Lower{a} & \Lower{b} &  \multicolumn{2}{c}{Koopmans} & \multicolumn{2}{c}{Modified Koopmans} & \multicolumn{2}{c}{DEA-EOM-pCCD(3p,1h)}  \\ \cline{6-11}
 & & & & & HF & pCCD & HF & pCCD & HF & pCCD  \\ 
He
  & cc-pVDZ  &  $\ce{^{1}S_{0}}$  & ${2s_a}$  & ${2s_{\bar{b}}}$ & 96.88   & 96.88 & 97.68  & 97.69 & 97.23  & 97.23  \\
  & cc-pVTZ  &   &  &  & 46.51  & 110.78 & 46.79  & 111.61 & 46.15  & 46.31 \\
  & cc-pVQZ  &  &  &  & 37.32  & 108.71 & 37.48  & 109.54 & 36.85  & 37.02 \\
 
 &\\

Be (ncore=0)
  & cc-pVDZ  & $\ce{^{3}P_{2}}$  &  ${2p_z}$ & ${2p_y}$  & 9.11  & 15.29 & 9.48 & 16.05 & 7.74 & 8.01   \\
   & cc-pVTZ   &  &  &  & 8.06  & 15.23 & 8.31  & 16.00  & 6.39  & --$^a$  \\
  & cc-pVQZ  & &  &  & 7.37   & 15.33 & 7.55   & 16.10  & 5.74  & 6.10   \\

  &\\

Ne
  & cc-pVDZ  & $\ce{^{1}S_{0}}$  & ${3s_a}$  & ${3s_{\bar{b}}}$  & 109.94  & 109.90 & 109.34  & 110.46   &102.53$^a$ & --$^a$ \\
   & cc-pVTZ  &   &  &  & 72.56  & 122.45  & 72.86  & 123.07  & 69.59$^a$ & --$^a$  \\
  & cc-pVQZ &  &  &  & 54.67  & 122.35 & 54.84   & 122.98  & --$^a$ & 52.83 \\

 &\\

Mg (ncore=1)
  & cc-pVDZ  & $\ce{^{3}P_{2}}$  &  ${3p_z}$ & ${3p_y}$  & 6.86 & 12.82 & 7.06   & 13.36   & 5.92 & 6.11    \\
  & cc-pVTZ  & &  &  & 5.90 & 12.66 & 6.02 & 13.20 & 4.81 & 5.04   \\
  & cc-pVQZ  &  &  &  & 4.87   & 9.21 & 4.92 & 9.48 & 3.89 & 4.12   \\
 
 &\\

Ar
  & cc-pVDZ  & $\ce{^{1}S_{0}}$  & ${4s_a}$  & ${4s_{\bar{b}}}$ & 53.42   & 53.46 & 53.58 & 53.65 & --$^a$ & --$^a$    \\
   & cc-pVTZ  & &  &  & 37.82   & 65.39 & 37.89  & 65.60 & --$^a$ & 35.22 \\
  & cc-pVQZ & &  &  & 27.71 & 66.83 & 27.74 & 67.04 & --$^a$ & 26.50    \\
 
 &\\

Ca (ncore=5)
  & cc-pVDZ  & $\ce{^{3}F_{2}}$  &  ${3d_{z^2}}$ & ${3d_{xz}}$  & 4.87  & 8.85  &  5.05   & 9.32  & 3.68 &  3.83   \\
  & cc-pVTZ  & &  &  & 4.64 & 9.50 & 4.79 & 9.73  & 2.78  & 3.02    \\
  & cc-pVQZ  & &  &  & 3.40 & 26.69 & 3.42 & 26.79 & 2.78 & 3.03   \\
  
 &\\

Zn
  & cc-pVDZ  & $\ce{^{3}P_{2}}$  &  ${4p_z}$ & ${4p_y}$  & 7.63 & 15.16  & 7.79 & 15.72 & 6.49   & 6.72    \\
   & cc-pVTZ &  &  &  & 7.56  & 15.36 & 7.71   & 15.92 & 6.35 & 6.60   \\
  & cc-pVQZ  &  &  &  & 7.65 & 16.94 & 7.76   & 17.51 & --$^a$ & --$^a$   \\
  
  &\\

Kr
  & cc-pVDZ  & $\ce{^{1}S_{0}}$  & ${5s_a}$  & ${5s_{\bar{b}}}$ & 49.59 & 49.65 & 49.68   & 49.74  & 44.53$^a$   & 44.56  \\
   & cc-pVTZ  & &  &  & 29.40 & 55.93 & 29.45  & 56.08 & 27.96$^a$   & --$^a$   \\
  & cc-pVQZ  &  &  &  & 19.85  & 56.89 & 19.87  & 57.05  & --$^a$ & --$^a$ \\
 
 &\\

 \end{tabular}
 \end{ruledtabular}

  \footnotetext[1]{1The DEA calculations for Be, Ne, Ar, Zn, and Kr atoms did not converge due to numerical difficulties. Convergence issues with the Davidson algorithm led to imaginary eigenvalues, resulting in calculations converged only to \(1 \times 10^{-3}\).}
  }
\end{table*}

\subsubsection{Molecules}
By evaluating DIPs of acceptor molecules (refer to Table~S7 of the SI), we see that Koopmans(HF) and modified Koopmans(HF) generally align with the results obtained from the DIP-EOM-pCCD method. 
However, when using the natural pCCD orbitals, the performance of both methods worsens, resulting in more significant deviations compared to DIP-EOM-pCCD. 
Since experimental data for DIPs of these molecules is not available, a comprehensive comparison cannot be made. 
An examination of the double electron affinities (DEAs) for molecular structures (see Table~S8 in the SI) reveals that the values predicted by both the Koopmans(HF) and modified Koopmans(HF) methods are consistently lower than those obtained using pCCD-optimized natural orbitals.
Unfortunately, we do not have any experimental data available for a direct comparison, which limits our ability to assess these computational methods' accuracy effectively.
{Yet, comparing Koopmans-based pair orbital energies (approximations to DIPs/DEAs) with the DIP/DEA-EOM results for
larger molecules is not straightforward.
While the Koopmans-based DIPs/DEAs correspond to one leading
electronic configuration, the corresponding DIP/DEA-EOM have non-negligible contributions from many other
orbitals.
That makes the one-to-one comparisons very difficult.
}

\section{Conclusions}\label{sec:conclusions}
In this work, we have developed a hierarchy of simple and cost-effective Koopmans-type models for computing orbital energies and pair orbital energies to be applicable in large-scale quantum chemical calculations. 
From these quantities, we can approximate IPs, EAs, DIPs, DEAs, and optical gaps, which are indispensable in the search for new building blocks of organic devices.  

Our modified Koopmans' approach incorporating some electron correlation effects from the pCCD model does not differ substantially from the original Koopmans formulation. 
The Koopmans and modified Koopmans' approaches can take advantage of pCCD-optimized natural orbitals leading to a more balanced description of occupied and virtual orbitals than the canonical ones. 
To that end, Koopmans(pCCD) and modified Koopmans(pCCD) flavors offer a valuable tool for predicting (a first estimation of) HOMO-LUMO gaps {in large molecular systems.}
They are computed at a small fraction of the cost of orbital optimization within the pCCD ansatz.
Due to their low computational cost (o$^2$v$^2$), the Koopmans(pCCD) approach and its modified variant represent {potential} models for screening novel organic compounds with application focus to organic electronics. 
\section{Supplementary Material}
See the supplementary material for additional information regarding additional information on (D)IP/(D)EA-EOM-CC methods, statistical analysis errors, and electron affinities and ionization potentials of acceptor molecules. 
\section{Acknowledgment}\label{sec:acknowledgement}
S.~J., S.~A., and P.~T.~acknowledge financial support from the SONATA BIS research grant from the National Science Centre, Poland (Grant No. 2021/42/E/ST4/00302). 
Funded/Co-funded by the European Union (ERC, DRESSED-pCCD, 101077420).
Views and opinions expressed are, however, those of the author(s) only and do not necessarily reflect those of the European Union or the European Research Council. Neither the European Union nor the granting authority can be held responsible for them.  

\bibliography{rsc} 
\bibliographystyle{rsc} 

\end{document}